\documentclass[12pt,preprint]{aastex}

\shorttitle{Effects of a Local Interstellar Magnetic Field}
\shortauthors{Opher, Stone and Liewer}
\begin{document}

\title{Effects of a Local Interstellar Magnetic Field on Voyager 1 and 2 Observations}

\author{Merav Opher\altaffilmark{1}}
\affil{George Mason University, 4400 University Drive, Fairfax, VA 22030}
\email{mopher@physics.gmu.edu}

\author{Edward C. Stone\altaffilmark{2}}
\affil{California Institute of Technology, Pasadena, CA 91125} 
\email{ecs@srl.caltech.edu}
 
\author{Paulett C. Liewer\altaffilmark{3}}
\affil{Jet Propulsion Laboratory, California Institute of Technology, Pasadena, CA 91109}
\email{paulett.liewer@jpl.nasa.gov}

\begin{abstract}
We show that that an interstellar magnetic field can produce a north/south asymmetry in solar wind termination shock. Using Voyager 1 and 2 measurements, we suggest that the angle $\alpha$ between the interstellar wind velocity and magnetic field is $30^{\circ} < \alpha < 60^{\circ}$.  The distortion of the shock is such that termination shock particles could stream outward along the spiral interplanetary magnetic field connecting Voyager 1 to the shock when the spacecraft was within  $\sim 2~AU$ of the shock.  The shock distortion is larger in the southern hemisphere, and Voyager 2 could be connected to the shock when it is within $\sim 5~AU$ of the shock, but with particles from the shock streaming inward along the field.  Tighter constraints on the interstellar magnetic field should be possible when Voyager 2 crosses the shock in the next several years. 
\end{abstract}


\section{Introduction}
The motion of solar system through the interstellar medium with a velocity of  $25.5~km/s$ compresses the heliosphere, producing a comet-like shape with an extended tail.  The heliosphere is created by the supersonic solar wind which abruptly slows, forming a termination shock as it approaches contact with the interstellar medium at the heliopause. Beyond the heliopause, the interstellar wind contains neutral atoms, mainly hydrogen and helium, and ions that carry the frozen-in interstellar magnetic field. 

The twin Voyager spacecraft are probing the northern and southern hemispheres of the heliosphere.  On December 16, 2004, Voyager 1 crossed the termination shock at $94~AU$ and began exploring the heliosheath \citep{burlaga,decker,stone05}. Voyager 1 is now beyond $98~AU$ at $34.1^{\circ}$ in latitude and $178^{\circ}$ in longitude, while Voyager 2 is beyond $78~AU$ at $-26.2^{\circ}$ in latitude and $213^{\circ}$ in longitude (solar ecliptic coordinates).  In mid 2002, Voyager 1 began observing strong energetic beams of termination shock particles (TSPs), streaming outward along the spiral magnetic field upstream the shock. \citet{jokipii} and \citet{stone05} suggested that the upstream beaming resulted from a non-spherical shock with the spiral interplanetary magnetic field line crossing the shock (the source of TSPs) and returning to the supersonic solar wind before reaching Voyager 1. 

An interstellar magnetic field oriented obliquely to the interstellar velocity can produce a lateral or north-south asymmetry in the heliospheric shape\citep{pog96,ratk}. However, the direction and intensity of the local interstellar magnetic field are not well constrained.  Based on the polarization of light from nearby stars, \citet{frisch} suggested that the magnetic Þeld direction is parallel to the galactic plane (and directed toward $l \sim 70^{\circ}$).  The source distribution of the Voyager 3kHz radio emission is also parallel to that plane \citep{kur}. On the other hand, \citet{lall} found from resonantly scattered solar Lyman-$\alpha$ radiation that the flow direction of interstellar neutral hydrogen differs from that of helium by $4^{\circ}$ in a direction consistent with an interstellar magnetic field plane inclined $60^{\circ}$ from the galactic plane. This deflection has been demonstrated by model calculations \citep{izmo}. In this paper, we will assume the interstellar field is parallel to this plane, which we will refer to as the the H-deflection plane (HDP). There are few previous MHD models of the interaction of the solar and interstellar winds that include both the solar magnetic field and the interstellar magnetic field\citep{linde,pog,mcnutt,wash}.

In this paper we explore the non-spherical shape of the termination shock and the orientation of the interstellar magnetic field and show that observations from Voyager 1 and 2 may provide information on the inclination and strength of the interstellar magnetic field. 

\section{Results}
The model is based on the BATS-R-US code, a three-dimensional magnetohdyrodynamic (MHD) parallel, adaptive grid code developed by University of Michigan\citep{gombosi} and adapted by \citet{opher03,opher04} for the outer heliosphere. We used a grid ranging from $1.5~AU$ to $20~AU$ with an inner boundary at $30~AU$ and an outer boundary at $-1500~AU$ and $1500~AU$ in the $y$ and $z$ directions and at $-800~AU$ to $800~AU$ in the $x$ direction. The solar magnetic field axis was aligned with the solar rotation axis with a 26 days solar rotation period. The solar wind was taken as uniform $450~km/s$; only the ionized component was included. The  solar wind at the inner boundary was $n=7.8\times 10^{-3} cm^{-3}$, $T=1.6\times 10^{3}~K$, and a Parker spiral magnetic field of $2\mu G$ at the equator. For the interstellar wind, we used 
$n=0.07~cm^{-3}$ and $T=10^{4}~K$ \citep{frisch96}. 

We considered interstellar magnetic fields in the H-deflection plane ({\it HDP}) with different inclination angles  $\alpha$, where $\alpha$ is the angle between the interstellar magnetic field and interstellar wind velocity. The coordinate system has the interstellar velocity direction in the $+x$ direction and the $z$-axis as the solar rotation axis of the sun, with $y$ completing the right handed coordinate system. In this coordinate system, Voyager 1 is at $29.1^{\circ}$ in latitude 
and $213.4^{\circ}$ in longitude and Voyager 2 is at $-31.2^{\circ}$ and $178.4^{\circ}$ in longitude, which ignores the $7.25^{\circ}$ tilt of the solar equator with respect to the ecliptic plane. 

Figure 1 shows a case for $B_{ISM}$ perpendicular to the flow ($\alpha=90^{\circ}$). $B_{ISM}=1.8\mu G$ for the cases in Figures 1 through 5. Subsonic flow carries the spiral solar field (black lines) into the hot heliosheath region (red) downwind of the termination shock (yellow surface). Only segments of the field lines appear in the nose region where the localized inward distortion of the shock results in field lines that cross into and back out of the heliosheath. Further from the nose the solar magnetic field lines remain in the heliosheath as they spiral outward. 

Figure 2 shows the contours of the magnetic field strength in the {\it HDP} plane for $B_{ISM}$ with an inclination $\alpha=45^{\circ}$ and $B_{ISM,y}<0$. A large north-south asymmetry in the heliosheath is apparent as is the northward deflection of the heliospheric current sheet (HCS) in the heliosheath which is thicker in the north (see also \citet{pog}). The smaller north/south asymmetry in the termination shock will be discussed further below.

The spiral solar magnetic field lines connecting to Voyager 1 and 2 lie on cones with latitudes of $29.1^{\circ}$ and $-26.2^{\circ}$ and the intersections of those cones with the termination shock are shown in Figure 3. In both the northern and southern hemisphere, the cones intersect the surface of the termination shock closer to the equator near the nose where the shock is closer to the Sun. 

Figure 4 shows the intersection of the termination shock with the cones and the solar magnetic field lines for Voyager 1 and 2. The distortion of the shock is such that the shock is closer to the Sun counter clockwise from Voyager 1. It can be seen that $2~AU$ inside the shock Voyager 1 was connected to the shock along the field line in the direction toward the Sun, allowing termination shock particles (TSPs) to stream outward along the field as observed. The distortion is larger in the southern hemisphere resulting in field lines $5~AU$ from the shock connecting the shock to Voyager 2. Also, the distortion is such that the shock is closer to the Sun clockwise from Voyager 2, so the TSP streaming should be inward along the spiral magnetic field line, opposite to that for Voyager 1, as has been observed \citep{cummings,deckerb}.

The radial distance of the shock at Voyager 1 and 2 decreases significantly as the inclination angle $\alpha$ increases from $0^{\circ}$ to $60^{\circ}$ with the shock $7$ to $10~AU$ closer in the south at Voyager 2 than in the north at Voyager 1 (Figure 5a). The maximum distortion of the shock at Voyager 1 is $\sim 2.0~AU$ (Figure 5b). Although there were no direct indications of how far Voyager 1 was from the shock when upstream episodes of TSPs were observed, MHD models based on Voyager 2 solar wind pressure measurements \citep{rich} suggest that the distance was less than $3$ to $4~AU$. This is somewhat larger than the maximum distortion in Figure 5b, suggesting that $\alpha$ is likely in the range of $30^{\circ}$  to $60^{\circ}$, where the distortion is maximum. 

The maximum distortion is much larger in the southern hemisphere and strongly depends on $\alpha$, with a maximum distortion at the latitude of Voyager 2 of $4$ to $7~AU$ for $\alpha =45^{\circ}$ to $60^{\circ}$. Thus Voyager 2 should observe TSPs somewhat further upstream from the shock than did Voyager 1. Because of the strong dependence on $\alpha$, the distance that TSPs are observed upstream of the shock by Voyager 2 should constrain the angle of inclination of the interstellar magnetic field.

The locations of the termination shock and heliopause scaled from various recent models are shown in Table 1, where the model results have been normalized to $90~AU$, the estimated average location of the shock at the latitude of Voyager 1 as the shock moves inward and outward over the solar cycle. Some conclusions that can be drawn from Table 1 are: a) For the models including an interplanetary magnetic field, the termination shock ranges between $76$ to $82~AU$ for Voyager 2; b) There is somewhat less North-South asymmetry with a weaker $B_{ISM}$; c) The heliosheath thickness on Voyager 1 is about $55$ to $59~AU$. For the model that has an interplanetary magnetic field and strong $B_{ISM}$, the thickness at Voyager 2 is $33$ to $45~AU$. Although, this is a preliminary calculation and does not incorporate a tilted current sheet, it does indicate the possibility of a significant north-south asymmetry in the heliosheath thickness; d) These models do not include neutrals that may affect the degree of asymmetry. However, works such as \citet{linde98} that included neutrals find similar north-south asymmetry with a ratio of $\sim 1.13$ for the north/south distance to the termination shock and $1.44$ for the ratio for the heliopause distances, similar to models without neutrals. \citet{izmo} using a kinetic treatment with the $B_{ISM}$ in the {\it HDP} plane, found a smaller north-south asymmetry. Their treatment, however, did not include the solar magnetic field. 

\section{Conclusions and Predictions}
This model calculation indicates that an interstellar magnetic field 
$(B\sim 2\mu G)$ in the H-deflection {\it HDP} plane can distort the heliosphere in a manner consistent with the streaming observed by Voyager 1 and Voyager 2. The model also indicates that such a  distortion will result in a significant north/south asymmetry in the distance to the shock and the thickness of heliosheath. It is reasonable to expect that Voyager 2 will encounter the shock in the next several years, providing additional information about the strength and inclination of the local interstellar magnetic field.

\acknowledgments
The authors would like to thanks the use of Columbia cluster NASA Ames. Part of this work is the result of research performed at the Jet Propulsion Laboratory of the California Institute of Technology under a contract with the National Aeronautics and Space Administration.

\clearpage

\begin{figure}
\includegraphics[width=5in]{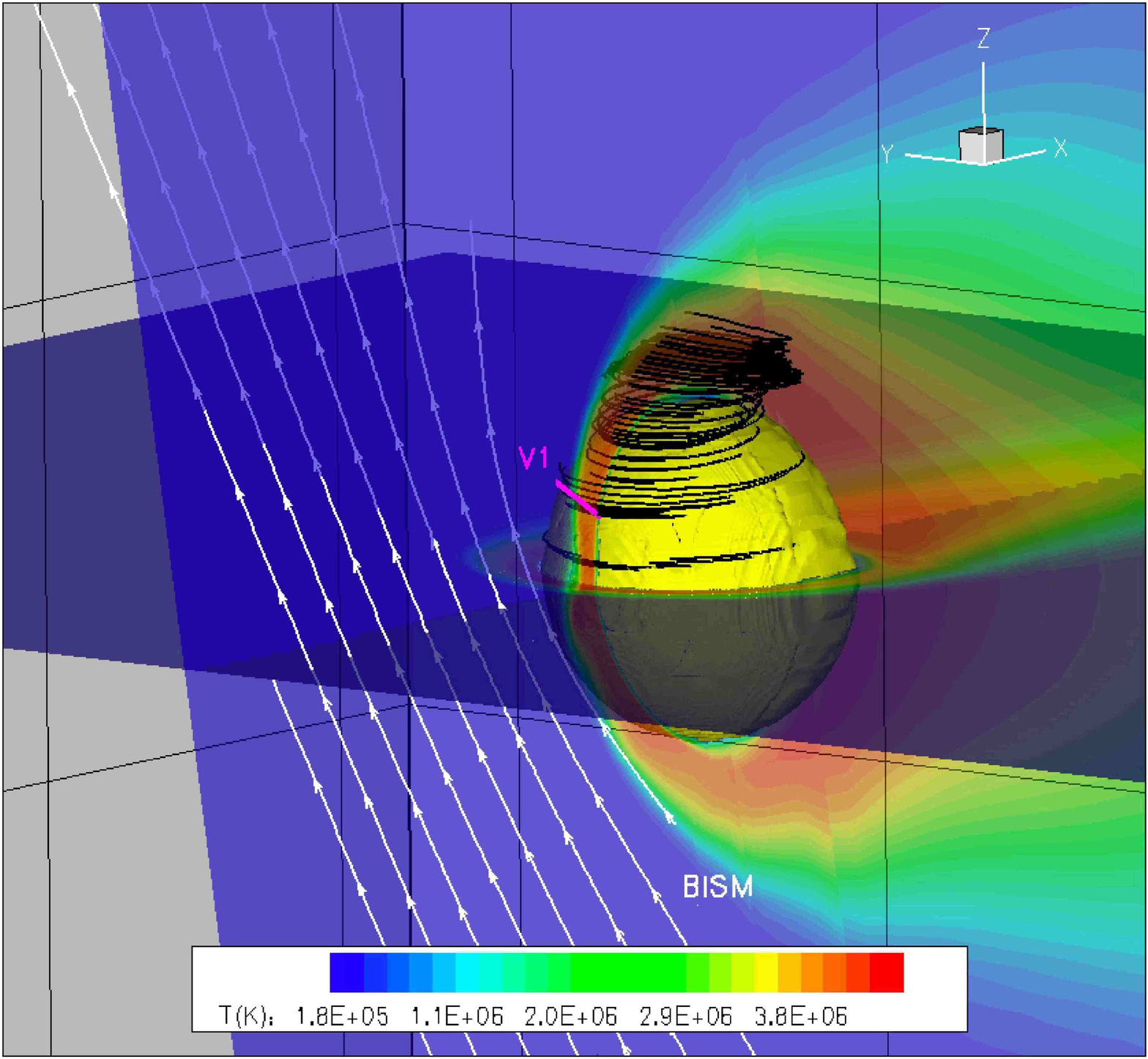} 
\caption{A model with $B_{ISM}$ in the {\it HDP} plane and ${\alpha}=90^{\circ}$ and $B_{ISM,y}<0$. The 3D iso-surface of the termination shock is denoted (yellow). The white lines are the interstellar magnetic field streamlines in the {\it HDP} plane and the black lines are the solar magnetic field lines in the heliosheath. The trajectory of Voyager 1 is denoted (magenta). The contours denote temperatures in the plane containing Voyager 1 and the $x$-axis.} 
\label{fig1}
\end{figure}

\clearpage

\begin{figure}
\includegraphics[width=5in]{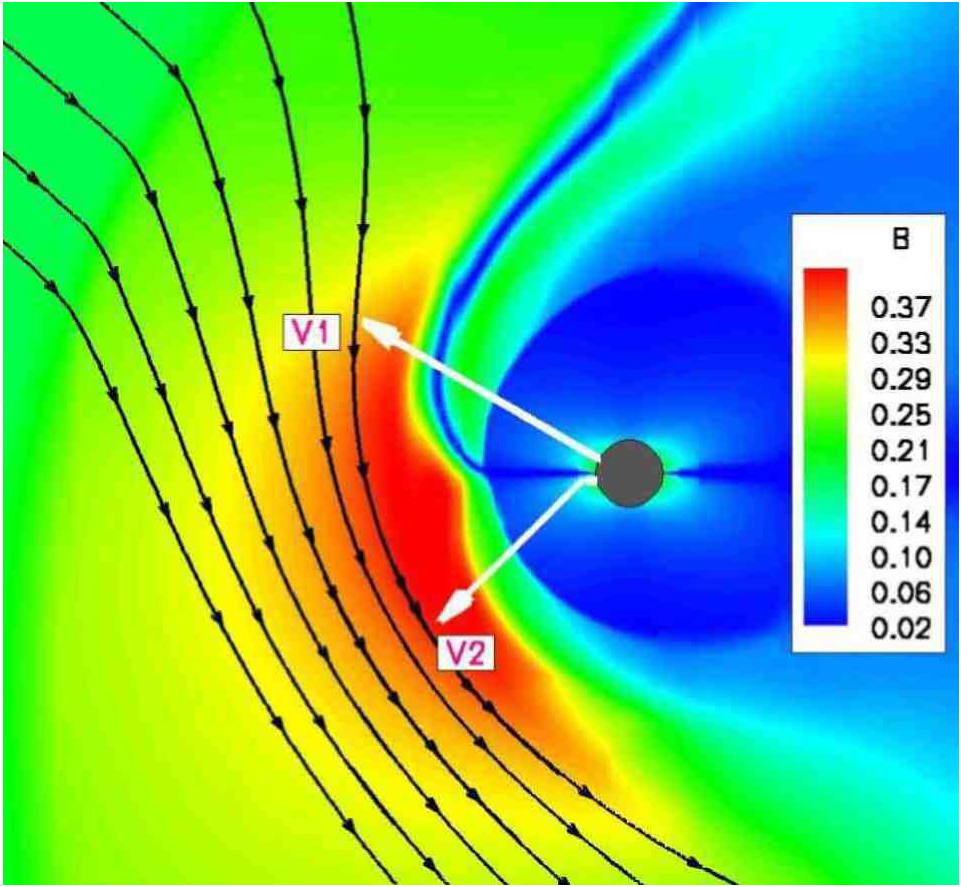} 
\caption{Contours of magnetic field strength $B(nT)$ in the  {\it HDP} plane with ${\alpha}=45^{\circ}$ and $B_{ISM,y}<0$. The black lines are the interstellar magnetic field and the white arrows denote the trajectories of Voyager 1 and Voyager 2. The heliospheric current sheet (deep blue) is deflected northward in the heliosheath.} 
\label{fig2}
\end{figure}

\clearpage

\begin{figure}
\includegraphics[width=5in]{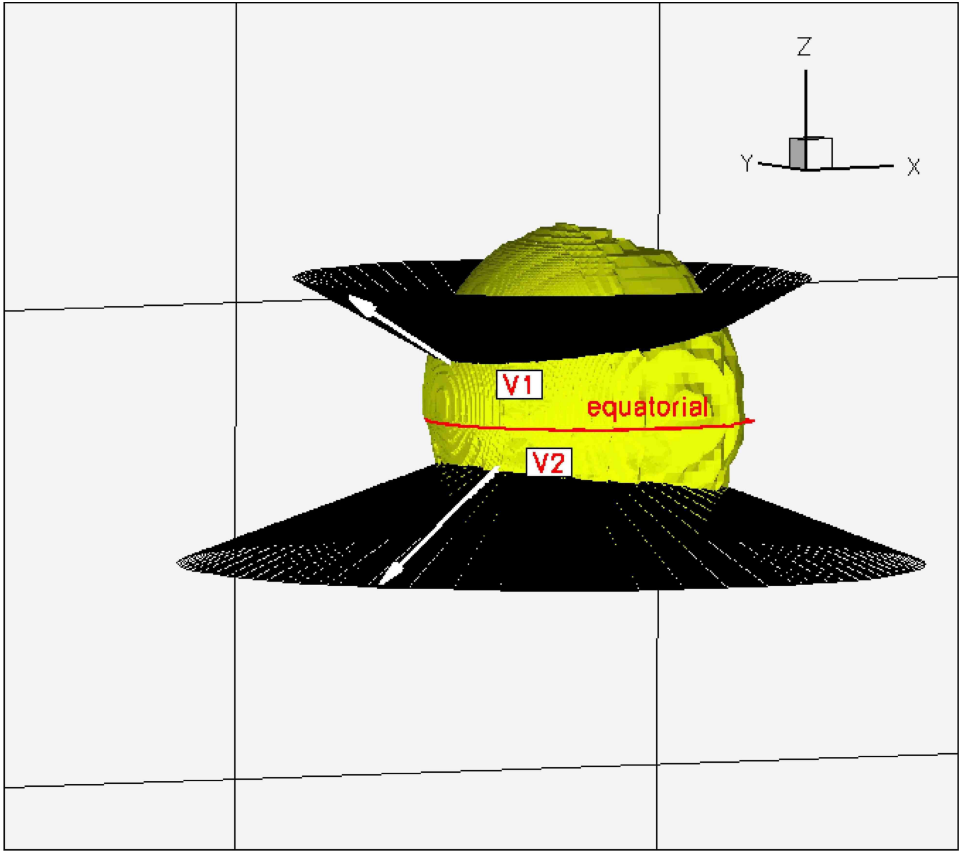}
\caption{The intersection of the termination shock surface (yellow) with the cone of interplanetary field line connecting to Voyager 1 and Voyager 2 (black lines). The trajectories of Voyager 1 and Voyager 2 are denoted (white arrows) as well as the solar equatorial plane (red line). $B_{ISM}$ is in the {\it HDP} plane with $\alpha=45^{\circ}$ and $B_{ISM,y}<0$.} 
\label{fig3}
\end{figure}

\clearpage

\begin{figure}
\includegraphics[width=3in]{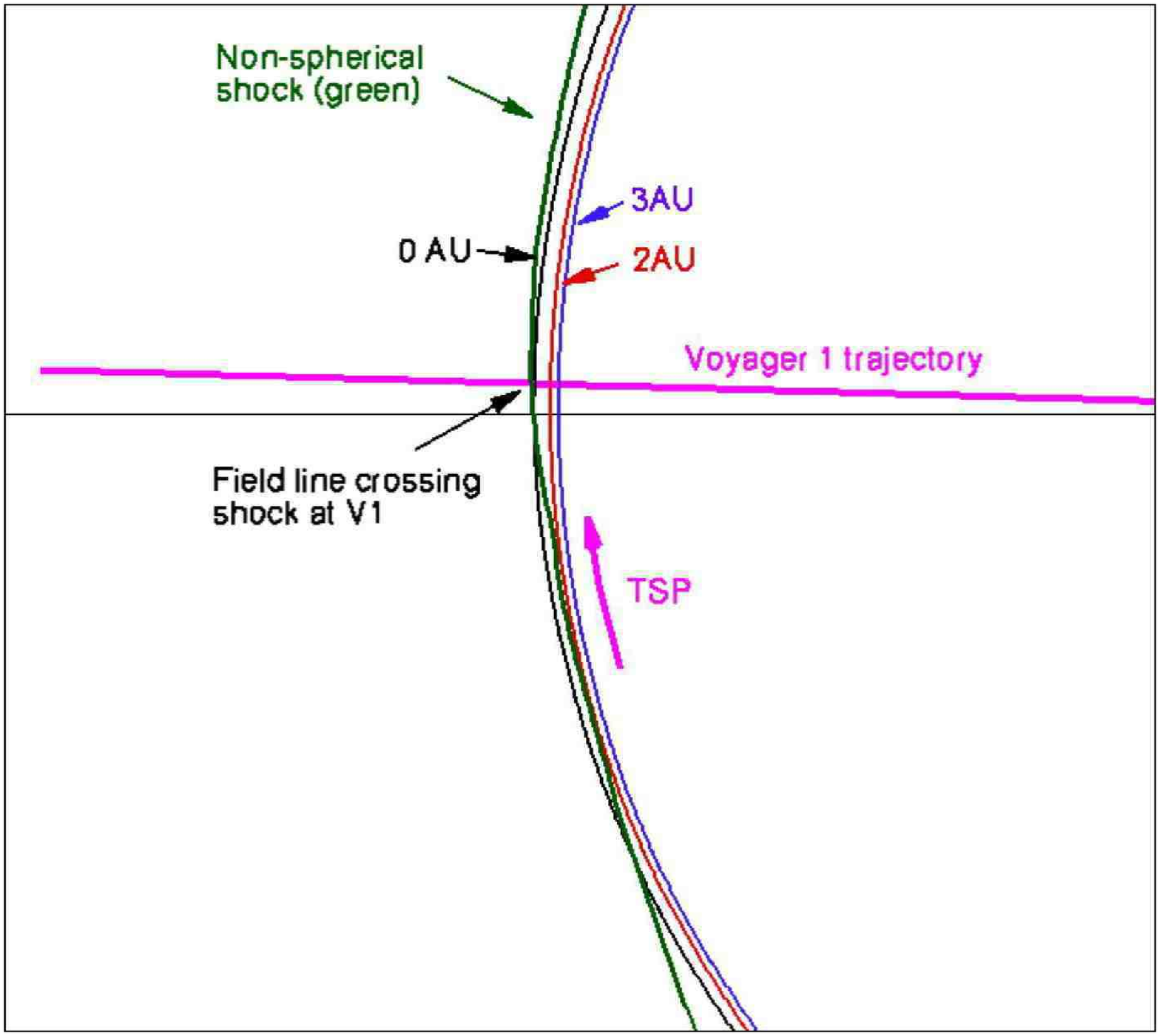} 
\includegraphics[width=3in]{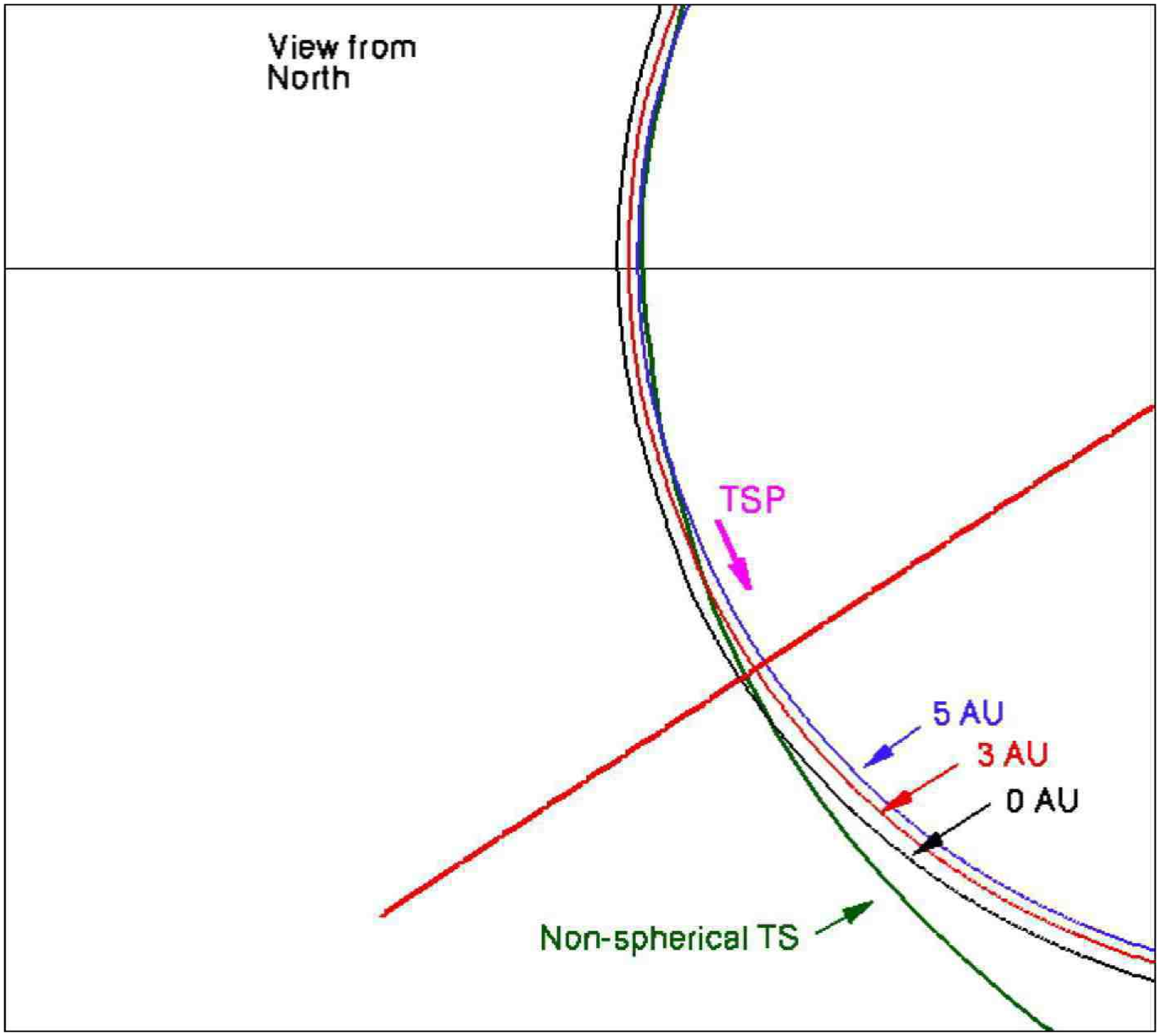} 
\caption{Expanded views from above of the field line cones in Figure 4. The plane of the plot is the surface of the interplanetary field line cones, the horizontal line is the intersection of conical surface with the $x-z$ plane, and the green line indicates the intersection of the termination shock with that conical surface. a) Spiral magnetic field lines on the Voyager 1 cone are shown; the field line intersecting the shock where Voyager 1 crosses the shock is labeled $0~AU$ (black) with red and blue indicating, respectively, magnetic field lines $2.0~AU$ and $3.0~AU$ upwind from the $0~AU$ line. The red arrow indicates the streaming direction of the termination shock particles from the shock along the field line to Voyager 1. b) Similar plot for Voyager 2, showing field lines $3.0$ and $5.0~AU$ upwind of the $0~AU$ line. Note that in both views the interplanetary magnetic field spirals outward with distance increasing clockwise.}
\label{fig4}
\end{figure}

\clearpage

\begin{figure}
\includegraphics[width=4in]{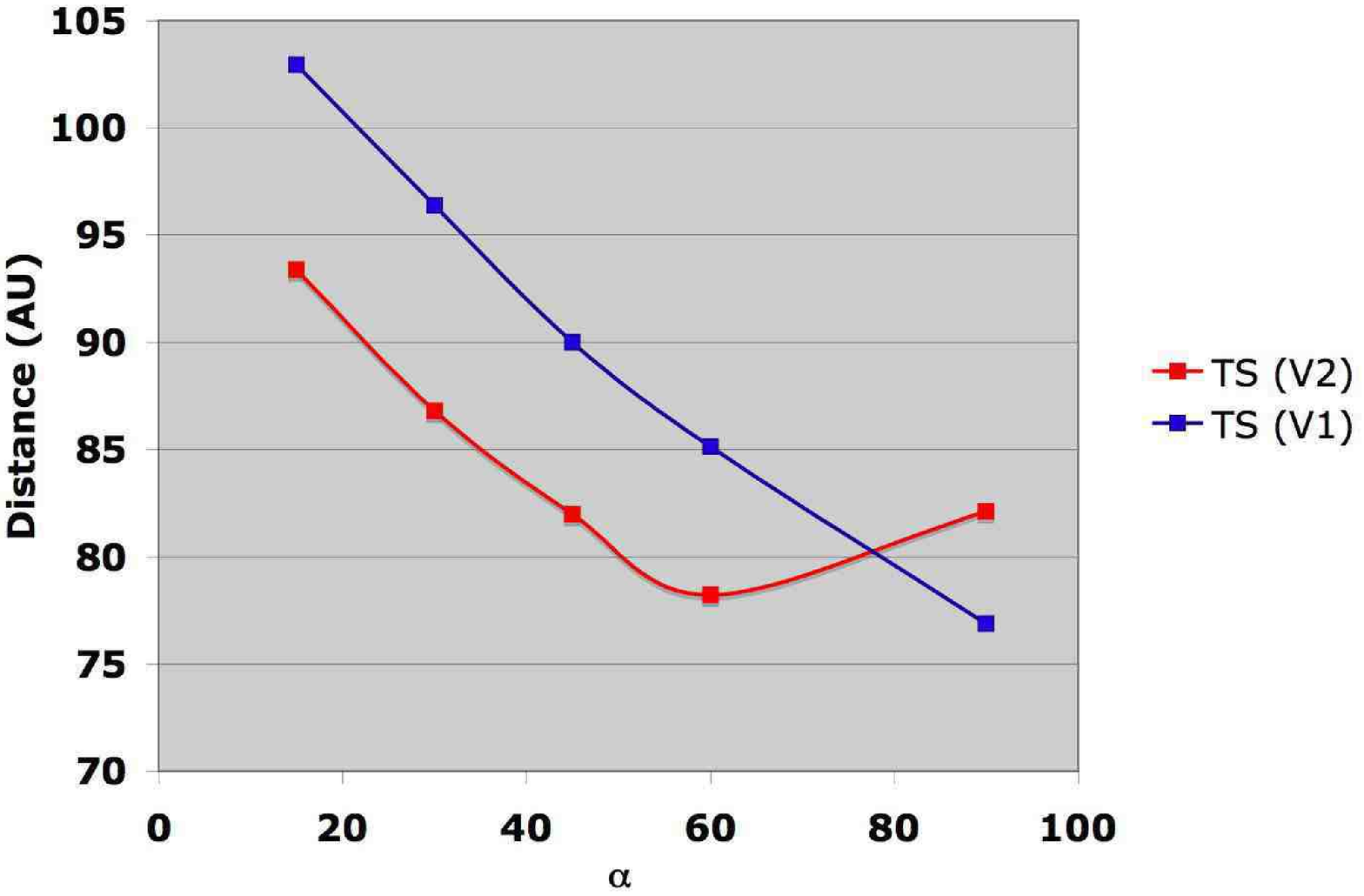} 
\includegraphics[width=4in]{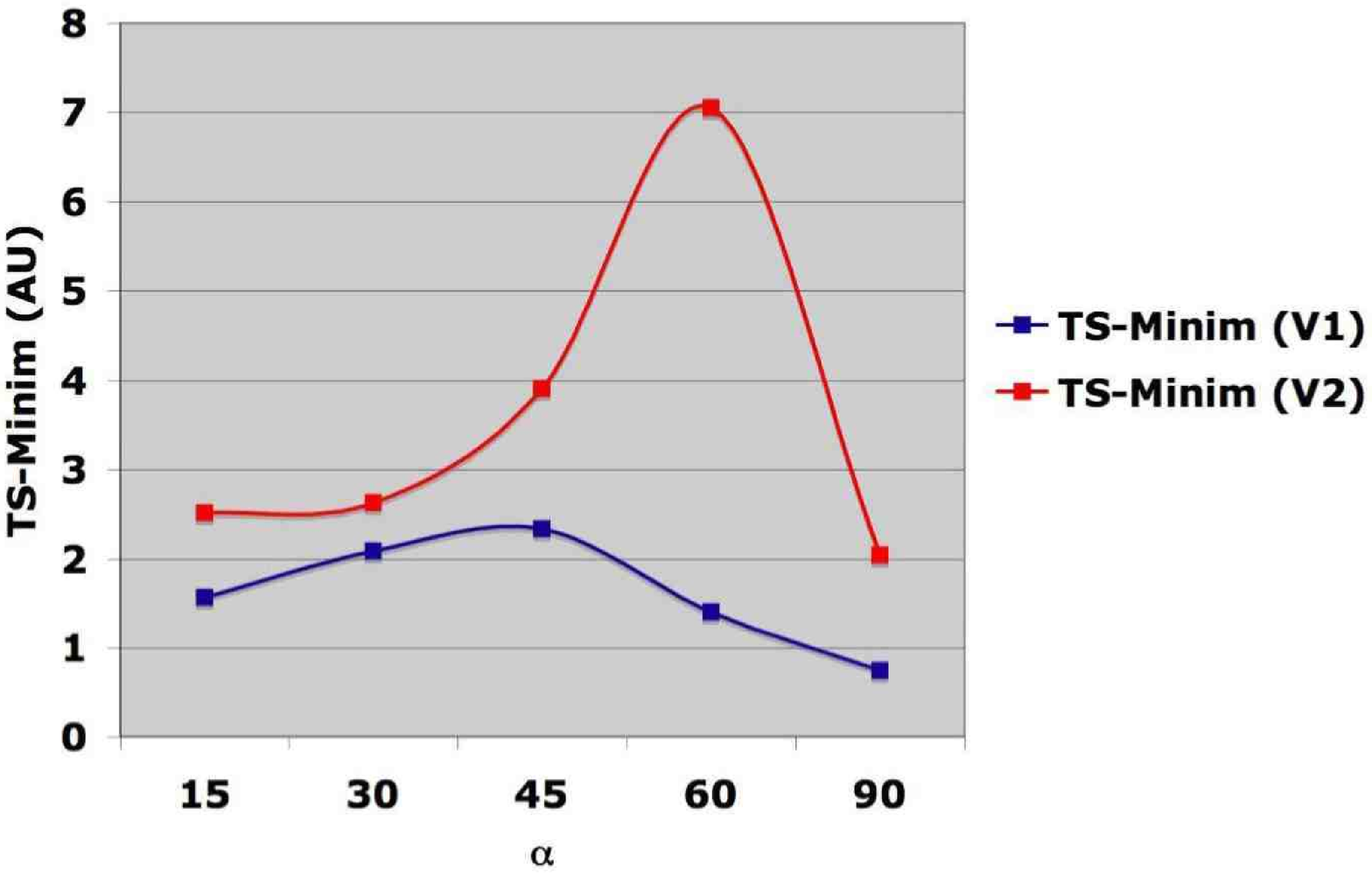} 
\caption{a) Distance to the termination shock at Voyager 1 (blue) and Voyager 2 (red) latitudes as a function of inclination ${\alpha}$ of the interstellar magnetic field. b) The difference between the distance to the shock at Voyager 1 and 2 longitude and the minimum distance to the shock as a function of ${\alpha}$ for Voyager 1 (blue) and for Voyager 2 (red). To match the estimated average shock location at Voyager 1, the shock locations have been normalized to $90~AU$ for $\alpha=45^{\circ}$. In addition to these results for $B_{ISM}= 1.8\mu G$, a model was also run for $2.5\mu G$ and is summarized in Table 1 along with locations determined from other recent model results.}
\label{fig5}
\end{figure}

\clearpage

\begin{deluxetable}{lccccccccccccc}
\tabletypesize{\scriptsize}
\tablecaption{Comparison of models. $B(\mu G)$ is the magnitude of $B_{ISM}$ with inclination angle $\alpha$; $P$ is the galactic (G) or the {\it HDP} (H) plane of $B_{ISM}$;$n$ indicates if neutrals are included or not; $B_{SW}$ indicates if interplanetary magnetic field is included; TS, the distance to the TS at Voyager 1 (V1) and Voyager 2
(V2) normalized to the estimated average distance measured at Voyager 1; HP the distance to the HP at V1 and V2; HS, the thickness of the heliosheath at V1 and V2; 
(TS), the ratio of the TS distances at V1 and V2: R(HS), the ratio of the heliosheath thicknesses. The models are: a) \citet{pog} (POG); b) 
\citet{izmo} (IZM); c) Our work (OSL); with 
$B_{ISW} =1.8\mu G$, $2.5 \mu G$ and a case with no solar magnetic field (OSL-Hydro); and d) \citet{linde98} (LI) with $B_{ISM} =1.5 \mu G$.}
\tablewidth{0pt}
\tablehead{
 & \colhead{B}  & \colhead{P} &  \colhead{n} & \colhead{$B_{SW}$} &\colhead{$\alpha$} & \colhead{TS} & \colhead{HP} & \colhead{TS} &\colhead{HP} & \colhead{HS} & \colhead{HS} & \colhead{R} & \colhead{R}\\
  & & & & & & \colhead{(V1)} & \colhead{(V1)} & \colhead{(V2)} & \colhead{(V2)} & \colhead{(V1)} &  \colhead{(V2)} & \colhead{(TS)} & \colhead{(HP)}\\
 }
 \startdata
{POG} & {2.4} & G &  N & Y & {45} & {90} & {145} & {76} & {109} & {55} & {33} & {1.18} & {1.33} \\
{IZM} & {2.4} &  H & Y & N & {45}  &  {90}  & {146} &  {91}  &  {139}  &  {56} &  {48} &  {0.99}  &  { 1.05}  \\
{OSL} &  {2.5}  &  H & N & Y & {45}  &  {90}  & {149} &  {79}  &  {124}  &  {59}  &  {45}  &  {1.13}  &  {1.20}  \\
{OSL} &  {1.8}  & H & N & Y& {45}  & {90}  & {145} & {82}  &  {132}  &  {55}  &  {50}  &  {1.10}  &  {1.10}  \\
{LI} & {1.5} & G & Y & Y &{71} & {90} & {157}  & {81}  & {130} & {67}  & {49}  &  {1.11}  &  {1.21} \\
{OSL-Hydro} & {1.8}  &  H & N & N & {45} &  {90}  & {130} & {87} & {127}  & {40}  & {40}  &  {1.03}  &  {1.02}  \\
 \enddata
\end{deluxetable}

\end{document}